\documentclass[a4paper,12pt]{article}

\usepackage[utf8]{inputenc}
\usepackage{makeidx}  
\usepackage{graphicx}
\usepackage[utf8]{inputenc}
\usepackage{tikz}
\usetikzlibrary{shapes,arrows}
\usepackage[framemethod=TikZ]{mdframed}
\usepackage{amsfonts}
\usepackage{amsmath}
\newcommand{\Mod}[1]{\ (\text{mod}\ #1)}
\usepackage{rotating}
\usepackage{multirow}
\DeclareMathAlphabet{\mathpzc}{OT1}{pzc}{m}{it}
\usepackage{amssymb}
\usepackage{latexsym}
\usepackage{filecontents}
\usepackage{url}

\title{Extended-Alphabet Finite-Context Models}

\author{João M. Carvalho \and Susana Brás \and Diogo Pratas \and Jacqueline Ferreira \and Sandra C. Soares \and Armando J. Pinho}

\begin{document}
	\maketitle

	\begin{abstract}
		The Normalized Relative Compression (NRC) is a recent dissimilarity measure, related to the Kolmogorov Complexity. It has been successfully used in different applications, like DNA sequences, images or even ECG (electrocardiographic) signal. It uses a compressor that compresses a target string using exclusively the information contained in a reference string. One possible approach is to use finite-context models (FCMs) to represent the strings. 
		
		A finite-context model calculates the probability distribution of the next symbol, given the previous $k$ symbols. In this paper, we introduce a generalization of the FCMs, called extended-alphabet finite-context models (xaFCM), that calculates the probability of occurrence of the next $d$ symbols, given the previous $k$ symbols.
		
		We perform experiments on two different sample applications using the xaFCMs and the NRC measure: ECG biometric identification, using a publicly available database; estimation of the similarity between DNA sequences of two different, but related, species -- chromosome by chromosome.
		
		In both applications, we compare the results against those obtained by the FCMs. The results show that the xaFCMs use less memory and computational time to achieve the same or, in some cases, even more accurate results.
		
	\end{abstract}

	\newpage

\section{Introduction}

Data compression models have been used to address several data mining and machine learning problems, usually by means of a formalization in terms of the information content of a string or of the information distance between strings \cite{Pinho2011a, Pratas2014a, Coutinho2015, Pinho2016a, Pratas2018}. This approach relies on solid foundations of the concept of algorithmic entropy and, because of its non-computability, approximations provided by data compression algorithms \cite{Li2004}. 

A finite-context model (\textrm{FCM}) calculates the probability distribution of the next symbol, given the previous $k$ symbols. In this work, we propose an extension of the \textrm{FCM}s, which we call extended-alphabet finite-context models (\textrm{xaFCM}). Usually, these models provide better compression ratios, leading to better results for some applications, especially when using small alphabet sizes -- and also by performing much less computations. We show this in practice for the ECG biometric identification and DNA sequence similarity. The source code for the compressor was implemented using Python 3.5 and is publicly available under the GPL v3 license\footnote{\url{https://github.com/joaomrcarvalho/xafcm}}.

\subsection{Compression-based measures}
Compression-based distances are tightly related to the Kolmogorov notion of complexity, also known as algorithmic entropy. Let $x$ denote a binary string of finite length. Its \textbf{Kolmogorov complexity}, $K(x)$, is the length of the shortest binary program $x^*$ that computes $x$ in a universal Turing machine and halts. Therefore, $K(x) = |x^*|$, the length of $x^*$, represents the minimum number of bits from which $x$ can be computationally retrieved \cite{KolmogorovA}. 

The \textbf{Information Distance} (ID) and its normalized version, the \textbf{Normalized Information Distance} (NID), were proposed by Bennett \textit{et al.} almost two decades ago \cite{Bennett1998} and are defined in terms of the Kolmogorov complexity of the strings involved, as well as the complexity of one when the other is provided.

However, since the Kolmogorov Complexity of a string is not computable, an approximation (upper bound) for it can be used by means of a compressor. Let $C(x)$ be the number of bits used by a compressor to represent the string $x$. We will use a measure based on the notion of \textit{relative compression} \cite{Pinho2016a}, denoted by $C(x||y)$, which represents the compression of $x$ relatively to $y$. 
This measure obeys the following rules:

\begin{itemize}
	\item $C(x||y) \approx 0$ iff string $x$ can be built efficiently from $y$;
	\item $C(x||y) \approx |x|$ iff $K(x|y) \approx K(x)$.
\end{itemize}

Based on these rules, the \textbf{Normalized Relative Compression} (\textrm{NRC}) of the binary string $x$ given the binary string $y$, is defined as
\begin{equation}
\textrm{NRC}(x||y)  = \frac{C(x||y)}{|x|},
\end{equation}
where $|x|$ denotes the length of $x$.

A more general formula for the \textrm{NRC} of string $x$, given string $y$, where the strings $x$ and $y$ are sequences from an alphabet $\mathcal{A} = \{ s_1, s_2, \dots s_{|\mathcal{A}|}\}$, is given by
\begin{equation}
\textrm{NRC}(x||y)  = \frac{C(x||y)}{|x| \log_2{|\mathcal{A}|}}.
\end{equation}

\section{Extended-Alphabet Finite-Context Models}	

\subsection{Compressing using extended-alphabet finite-context models}\label{sub_sec_xaFCM}

Let $\mathcal{A} = \{ s_1, s_2, \dots s_{|\mathcal{A}|}\}$ be the alphabet that describes the objects of interest.
An extended-alphabet finite-context model (\textrm{xaFCM}) complies to the Markov property, i.e., it estimates the probability of the next sequence of $d > 0$ symbols of the information source (depth-$d$) using the $k > 0$ immediate past symbols (order-$k$ context).
Therefore, assuming that the $k$ past outcomes are given by $x_{n-k+1}^{n} = x_{n-k+1} \cdots x_{n}$, the probability estimates, $P(x_{n+1}^{n+d}|x_{n-k+1}^{n})$ are calculated using sequence counts that are accumulated, while the information source is processed,

\begin{equation}\label{eq_fcm2}
P(w|x_{n-k+1}^{n}) = \frac{ v(w|x_{n-k+1}^{n}) + \alpha }{  v(x_{n-k+1}^{n}) + \alpha|\mathcal{A}|^{d} } 
\end{equation}
where $\mathcal{A}^d = \{ w_1, w_2, \dots w_{|\mathcal{A}|},\ldots w_{|\mathcal{A}|^d}\}$ is an extension of alphabet $\mathcal{A}$ to $d$ dimensions, 
$v(w|x_{n-k+1}^{n})$ represents the number of times that, in the past, sequence $w \in {\mathcal{A}^d}$
was found having $x_{n-k+1}^{n}$ as the conditioning context and where

\begin{equation}
v(x_{n-k+1}^{n}) = \sum_{a \in \mathcal{A}^{d}} v(a|x_{n-k+1}^{n}) 
\end{equation}
denotes the total number of events that has occurred within context $x_{n-k+1}^{n}$. 

In order to avoid problems with ``shifting" of the data, the sequence counts are performed symbol by symbol, when learning a model from a string.

Parameter $\alpha$ allows controlling the transition from an estimator initially assuming a uniform distribution to a one progressively closer to the relative frequency estimator. 

The theoretical information content average provided by the $i$-th sequence of $d$ symbols from the original sequence $x$, is given by

\begin{equation}
- \log_2 P(X_i = t_i  | x_{id-k}^{id-1}) \text{ bits,} 
\end{equation}
where $t_i = x_{id}, x_{id+1} \cdots x_{(i+1)d-1}$.

\quad

After processing the first $n$ symbols of $x$, the total number of bits generated by an order-$k$ with depth-$d$ \textrm{xaFCM} is equal to

\begin{equation}
- \sum_{i=1}^{n/d} \log_2 P(t_i|x_{di-k}^{di-1}),
\end{equation}
where, for simplicity, we assume that $n \Mod d = 0$.

\quad

If we consider a \textrm{xaFCM} with depth $d = 1$, then it becomes a regular \textrm{FCM} with the same order $k$. In that sense, we can consider that a \textrm{FCM} is a particular case of a \textrm{xaFCM}.

\quad

An intuitive way of understanding how a \textrm{xaFCM} works is to think of it as a \textrm{FCM} which, for each context of length $k$, instead of counting the number of occurrences of symbols of $\mathcal{A}$, counts the occurrences of sequences $w \in \mathcal{A}^d$. In other words, for each sequence of length $k$ found, it counts the number of times each sequence of $d$ symbols appeared right after it.

Even though, when implemented, this might use more memory to represent the model, an advantage is that it is possible to compress a new sequence of length $m$, relatively to some previously constructed model, making only $m/d$ accesses to the model. This significantly reduces the time of computation, as we will show in the experimental results presented in Sections \ref{section_ecg} and \ref{section_dna}.

\quad

Since, for compressing the first $k$ symbols of a sequence, we do not have enough symbols to represent a context of length $k$, we always assume that the sequence is ``circular". For long sequences, specially using small contexts/depths, this should not make much difference in terms of compression, but as the contexts/depths increase, this might not be always the case.

\quad

Since the purpose for which we use these models is to provide an approximation for the number of bits that would be produced by a compressor based on them, whenever we use the word ``compression", in fact we are not performing the compression itself. For that, we would need to use an encoder, which would take more time to compute. It would also be needed to add some side information for the compressor to deal with the circular sequences -- but that goes out of scope for our goal.

\subsubsection{Example}

Let $x$ be the circular sequence \textit{AAABCC}. Using a regular \textrm{FCM} with $k = 2$ and $\alpha = 0.01$, we would build the model from Table~\ref{table_fcm_model_tradicional} to represent $x$.

\begin{table}[]
	\centering
	\caption{\textrm{FCM} representation of the sequence \textit{AAABCC}.}
	\label{table_fcm_model_tradicional}
	
	\begin{tabular}{|c|c|c|c|c|c|}
		\hline
		Context $c$ & $v(A|c)$ & $v(B|c)$ & $v(C|c)$& $v(c) = \sum_{a \in \mathcal{A}} v(a|c)$\\\hline	
		BC  & 0  & 0 & 1 & 1 \\\hline		
		CA  & 1  & 0 & 0 & 1 \\\hline	
		AB  & 0  & 0 & 1 & 1 \\\hline	
		CC  & 1  & 0 & 0 & 1 \\\hline
		AA  & 1  & 1 & 0 & 2 \\\hline
	\end{tabular}
\end{table}

It is easy to notice that this representation can be implemented using an hash-table of strings to arrays of integers with fixed size (alphabet size $+ 1$). However, we propose a different alternative, which consists of building a hash-table of hash-tables. The reason for doing so is that often the number of counts of symbols for each context is very sparse, which would be a waste of memory. To represent exactly the same model, we would build the structure presented in Table~\ref{table_fcm_model_proposed_hash}.

\begin{table}[]
	\centering
	\caption{Proposed \textrm{xaFCM} representation of the sequence \textit{AAABCC} (with $d=1$). Notice that this model has exactly the same information as the one in Table~\ref{table_fcm_model_tradicional}.}
	\label{table_fcm_model_proposed_hash}
	
	\begin{tabular}{|c|c|c|c|}
		\hline
		Context $c$ & & & \\\hline	
		BC  & C: 1 & Total: 1 & \\\hline		
		CA  & A: 1  & Total: 1 & \\\hline	
		AB  & C: 1 & Total: 1 & \\\hline	
		CC  & A: 1  & Total: 1 & \\\hline
		AA  & A: 1 & B: 1 & Total: 2\\\hline
	\end{tabular}
\end{table}

For compressing the sequence $x$, relatively to itself, we would need $C(x||x)$ bits, where

\begin{multline}\label{eq_fcm_example}
C(x||x) = C(A|CC) + C(A|CA) + C(A|AA) + C(B|AA) +\\+ C(C|AB) + C(C|BC)
\end{multline}

and,

\begin{multline}
C(A|CC) = C(A|CA) = C(C|AB) = C(C|BC) = \\= -\log_2{\frac{1+ 0.01}{1+3 \times 0.01}} = 0.0283
\end{multline}
and
\begin{equation}
C(A|AA) = C(B|AA) = -\log_2{\frac{2+ 0.01}{1+3 \times 0.01}} = 1.007,
\end{equation}
which means $C(x||x) = 2.1272$ or, in other words, it is possible to compress $x$ relatively to itself using just 2.1272 bits.

\quad

Using a xaFCM, also with $k = 2$ and $\alpha = 0.01$, but with $d=2$, we would build the model presented in Table~\ref{table_rfcm_model_proposed_hash} to represent $x$.

\begin{table}[]
	\centering
	\caption{Proposed \textrm{xaFCM} representation of the sequence \textit{AAABCC} (with $d=2$).}
	\label{table_rfcm_model_proposed_hash}
	
	\begin{tabular}{|c|c|c|c|}
		\hline
		Context $c$ & & & \\\hline	
		BC  & CA: 1 & Total: 1 & \\\hline		
		CA  & AA: 1  & Total: 1 & \\\hline	
		AB  & CC: 1 & Total: 1 & \\\hline	
		CC  & AA: 1  & Total: 1 & \\\hline
		AA  & AB: 1 & BC: 1 & Total: 2\\\hline
	\end{tabular}
\end{table}

Therefore,

\begin{equation}\label{eq_fcm_example}
C(x||x) = C(AA|CC) + C(AB|AA) + C(CC|AB)
\end{equation}
where,

\begin{multline}
C(AA|CC) = C(CC|AB) = \\= -\log_2{\frac{1+0.01}{1+3^2 \times 0.01}} = 0.110
\end{multline}
and
\begin{equation}
C(AB|AA) = -\log_2{\frac{1+0.01}{2+3^2 \times 0.01}} = 1.049
\end{equation}
which means $C(x||x) = 1.269$ or, in other words, using a \textrm{xaFCM} to represent the sequence $x$ it is possible to compress it relatively to itself using just 1.269 bits.

\quad

Calculating the \textrm{NRC} for both compressors we obtain:

\begin{itemize}
	\item \textbf{Using \textrm{FCM} --} $\textrm{NRC}(x||x) = \frac{2.1272}{6 \times \log_2{3}} = 0.224$;
	\item \textbf{Using \textrm{xaFCM} --} $\textrm{NRC}(x||x) = \frac{1.049}{6 \times \log_2{3}} = 0.110$.
\end{itemize}

Based on this example, we can infer that, at least for some cases, it is possible to obtain better compression ratios, using \textrm{xaFCM}s instead of traditional \textrm{FCM}s to represent a sequence.

\subsection{Parameter Selection}

\subsubsection{Selection of $\alpha$}

Since adjusting the $\alpha$ parameter might not be trivial, as it depends on the choice of $d$ as well as on the alphabet size. It is, however, possible to choose $\alpha$ based on a certain desired probability $p$ for a specific outcome.

In our experiments, in order to avoid having one more parameter to ``tweak", we are defining $\alpha$ automatically, in a way such that, if sequence $w \in \mathcal{A}^d$ was only found once after a certain context $c = x_{n-k+1}^{n} \in x$, and no other sequence $\in \mathcal{A}^d$ was found after that context $c$ (in other words, the total of that line, in the model, is 1), we want to be $90\%$ sure that the same situation happens when compressing a sequence relatively to the learned model. In other words, when we calculate the number of bits,

\begin{equation}
- \log_2 P(X_i = t_i  |c) 
\end{equation}
needed to compress sequence $t_i = x_{id}, x_{id+1} \cdots x_{id + d - 1}$, we want to choose an $\alpha$ such that

\begin{equation}
P(X_i = t_i  | c) = 0.9^d.
\end{equation}

But, since
\begin{equation}
P(w|c)  =	\frac{ v(w|c) + \alpha }{  v(c) + \alpha|\mathcal{A}|^d },
\end{equation}
where, we have chosen $c$ and $w$ such that $v(c) = v(w|c) = 1$. Therefore

\begin{subequations}
	
	\begin{equation}
	P(w|c) = 0.9^d \iff \frac{1 + \alpha}{1+\alpha |\mathcal{A}|^d}= 0.9^d.
	\end{equation} Since $\mathcal{A}^d$ is an extension of $\mathcal{A}$ to $d$ dimensions,
	\begin{equation}
	\frac{1 + \alpha}{1+\alpha |\mathcal{A}|^d} = 0.9^d.
	\end{equation}
\end{subequations}

Also, since both the alphabet size $\mathcal{A}$ and the depth $d$ are static parameters, it is easy to solve the equation and choose $\alpha$ in this way. It is also worth mentioning that there is always a possible solution for the equation, since the denominator of the fraction on the left is never equal to zero.

\subsubsection{Selection of $d$}

The parameter $d$ is an integer greater or equal to one. As mentioned in Section \ref{sub_sec_xaFCM}, when $d = 1$, we are using a \textrm{xaFCM} which is equivalent to a \textrm{FCM} of the same order $k$. Therefore, they both produce exactly the same number of bits.

As $d$ increases, so does the RAM needed to store the \textrm{xaFCM} model -- but there is not much of an impact (for $d = 11$ the increase in memory usage is about 10\%). The reason for the model complexity to only increase this is that the number of different ``leafs" in the hash-tables does not change with the choice of $d$ -- only the size of each string stored does.

Something to take into account when choosing $d$ is that, the greater the value of $d$, the harder it would be for an arithmetic encoder to complete its process. Since we only want to compute the \textrm{NRC}, we do not use an encoder. However, to avoid unrealistic results, we want to choose a $d$ that produces an alphabet size of, at most, the \textrm{MaximumValue}(integer)$-1$ (e.g. $2^{31} - 1$) symbols. For that reason, using an alphabet of size $6$, we can say that $1 \leq d \leq 11$.

Often, we are mostly interested in the time it takes to compress a new target sequence, given an already built model representing the reference sequence. With this application in mind, we can say for sure that the $d$ should be as big as possible, since, as mentioned before, less computations need to be done to compress a new target sequence and, therefore, much less time is needed. Results from real experiments can be seen in the next section.

\section{Application 1 -- ECG Biometric Identification}\label{section_ecg}

In previous works, we have addressed the topic of ECG based biometric identification using a measure of similarity related to the Kolmogorov complexity, called the Normalized Relative Compression (\textrm{NRC}). To attain the goal, we built finite-context models (\textrm{FCM}) to represent each individual \cite{Bras2015a, carvalhoibpria2017} -- a compression-based approach that has been shown successful for different pattern recognition applications \cite{Pinho2011b, Pratas2014a,  Pinho2016a}. Other recent works, also based on a compression approach, use the Ziv-Merhav cross parsing algorithm for attaining the same goal \cite{Coutinho2013, Coutinho2010}.

Compression-based approaches found in the literature for ECG biometric identification does not seem to take advantage of the fact that the ECG is a quasi-periodical time-series. Since our method uses a semi-fiducial approach (it only detects the R-peak), it is trivial to know where the repetition should happen and take advantage of that fact. From previous results \cite{carvalhorecpad2016}, we concluded that, when consecutive \textit{heartbeats}\footnote{For readability, by ``heartbeat" we mean the interval between two consecutive R-peaks.} present low levels of noise, their quantization is almost identical.
As a consequence of this, we consider that any sequence we analyze is a circular sequence \cite{Grossi2016}. From this result, it is possible to infer that, compressing the beginning of an heartbeat using the end of the same heartbeat, may be identical to compress it using the end of the previous heartbeat. This may not sound as an advantage, however, this fact allows us to use heartbeats that are not consecutive, when performing the identification of a participant.

Since the purpose of this paper is to introduce the method, and not to focus too much in the ECG signal, we do not explore this fact. However, it is already being taken it into account when building the algorithm (one of the arguments that the algorithm accepts as input is the length of the expected repetition -- i.e. for this application, how many symbols has one heartbeat), because it will be important for building a real system, as we expect more noise to be present and, therefore, some segments need to be discarded when performing the compression \cite{carvalhorecpad2016}.

\subsection{R-peak detection}	
The development of a robust automatic \textit{R-peak} detector is essential, but it is still a challenging task, due to irregular heart rates, various amplitude levels and \textit{QRS} morphologies, as well as all kinds of noise and artifacts \cite{Kathirvel2011}. 

We decided to use a \textit{semi-fiducial} method for segmenting the ECG signal and, since this was not the major focus of the work, we used a preexisting implementation to detect \textit{R-peaks}, based on the method proposed in \cite{Kathirvel2011}. The reason for using a semi-fiducial approach is that fiducial methods have a higher error of detection, while detecting the R-peaks is, nowadays, an almost trivial process \cite{Kathirvel2011}. The method used detects the \textit{R-peaks} by calculating the average points between the \textit{Q} and \textit{S} points (from the \textit{QRS-complexes}) -- this may not give the real local maxima corresponding to the \textit{R-peaks}, but it produces a very close point. 

For more information regarding the process used for detecting the R-peaks check \cite{Kathirvel2011}.  The process was already validated by its authors using the standard MIT-BIH arrhythmia database, achieving an average sensitivity of 99.94\% and a positive predictivity of 99.96\%. It uses bandpass filtering and differentiation operations, aiming to enhance the \textit{QRS} complexes and to reduce out-of-band noise. A nonlinear transformation is used to obtain a positive-valued feature signal, which includes large candidate peaks corresponding to the \textit{QRS} complex regions.

\subsection{Quantization}	

Data compression algorithms are symbolic in nature. Text and DNA sequences are well-known examples of symbolic sequences, with well-defined associated alphabets. Contrarily the ECG signal needs first to be transformed into symbols before data compression can be applied. 

In this work, we have relied on the \textrm{SAX} (Symbolic Aggregate ApproXimation) representation \cite{Lin2003} to transform the ECG into a symbolic time-series.

We consider that the signal is already discrete in the time domain, i.e., that it is already sampled. However, we perform re-sampling using the previously detected R-peaks.	

There is a fundamental trade-off to take into account while performing the choice of the \textit{alphabet size}: the quality produced versus the amount of data necessary to represent the sequence \cite{Reading2013}. We tested the experiments using alphabet sizes from 3 up to 20 symbols and using different numbers of symbols each R-R segment (per heartbeat), and found that a combination of using an alphabet size of 6 and 200 symbols per heartbeat produced a good balance between the complexity of the strings/models and the accuracies obtained for biometric identification, allowing us to proceed with the improvement of the compressors, while these parameters remained static. However, this result does not guarantee that the same will hold true for a different dataset or application, nor does it guarantee that these are the optimal parameters. Future work is needed to perform this choice in a more robust and automatic way.

\subsection{Experimental Results}

\begin{figure}[!htb]
	\centering
	
	\begin{minipage}[b]{0.4\textwidth}
		\includegraphics[width=\textwidth]{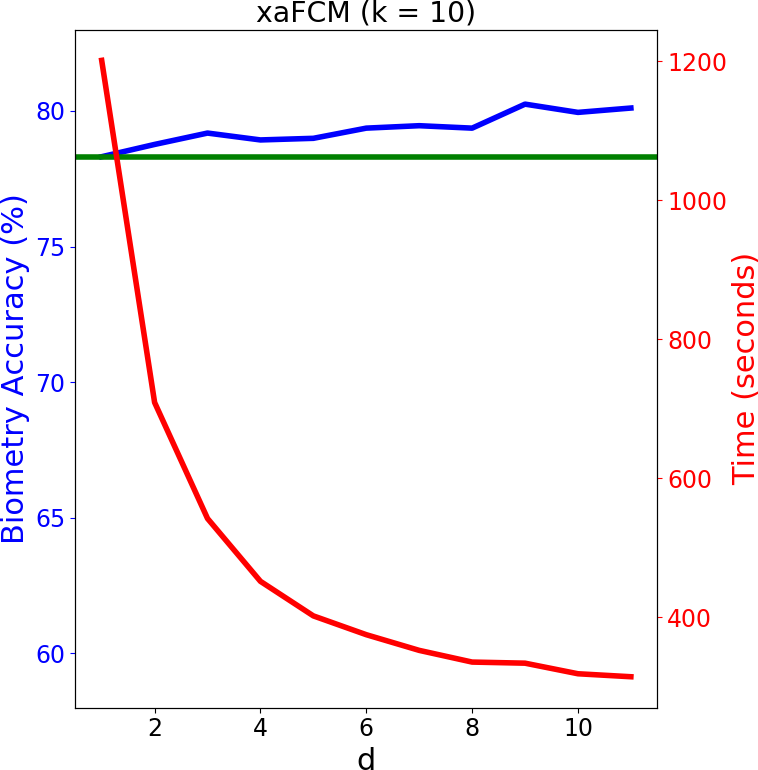}
	\end{minipage}
	\hfill
	\begin{minipage}[b]{0.4\textwidth}
		\includegraphics[width=\textwidth]{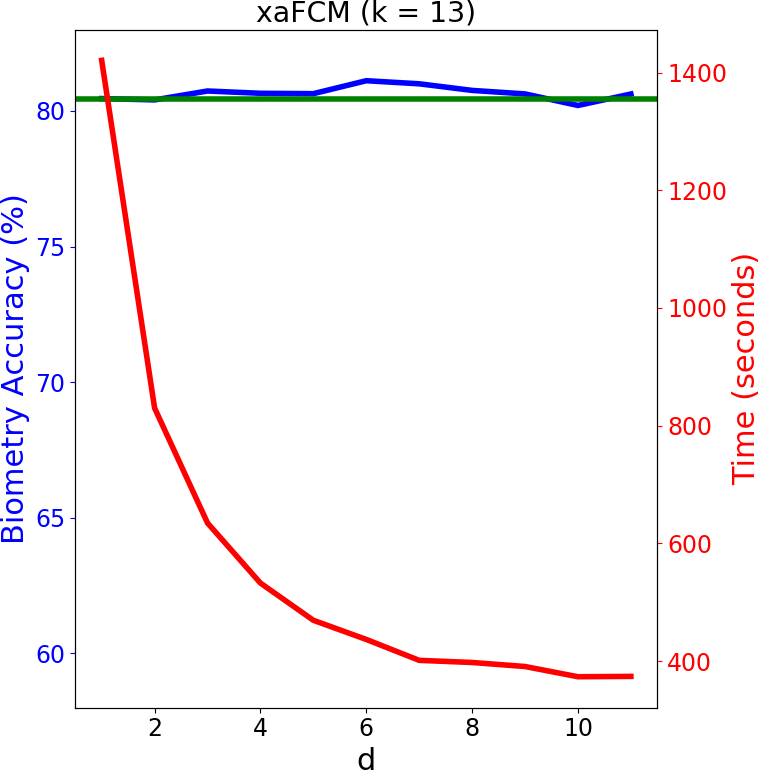}
	\end{minipage}
	
	\quad
	
	\begin{minipage}[b]{0.4\textwidth}
		\includegraphics[width=\textwidth]{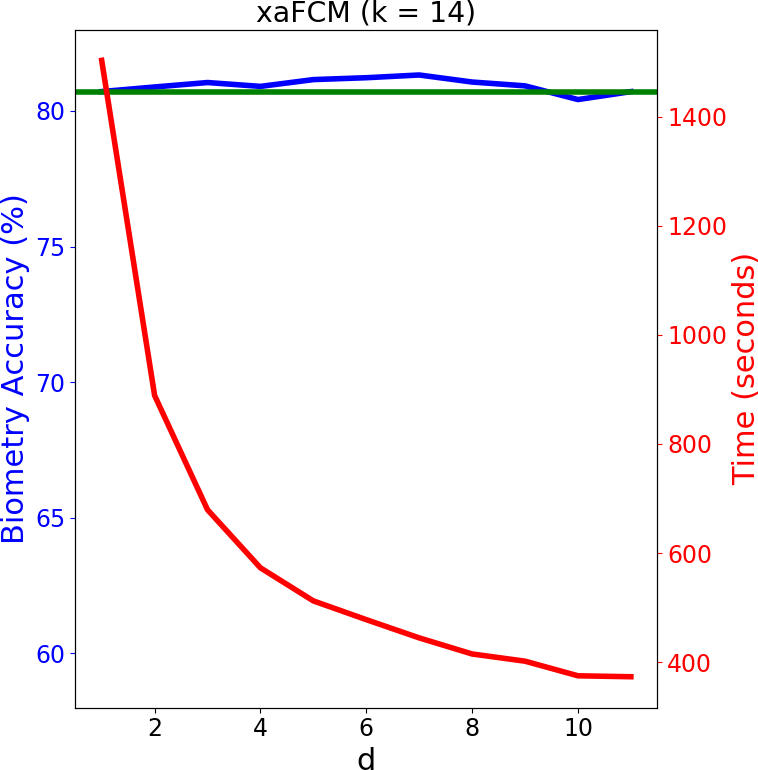}
	\end{minipage}
	\hfill
	\begin{minipage}[b]{0.4\textwidth}
		\includegraphics[width=\textwidth]{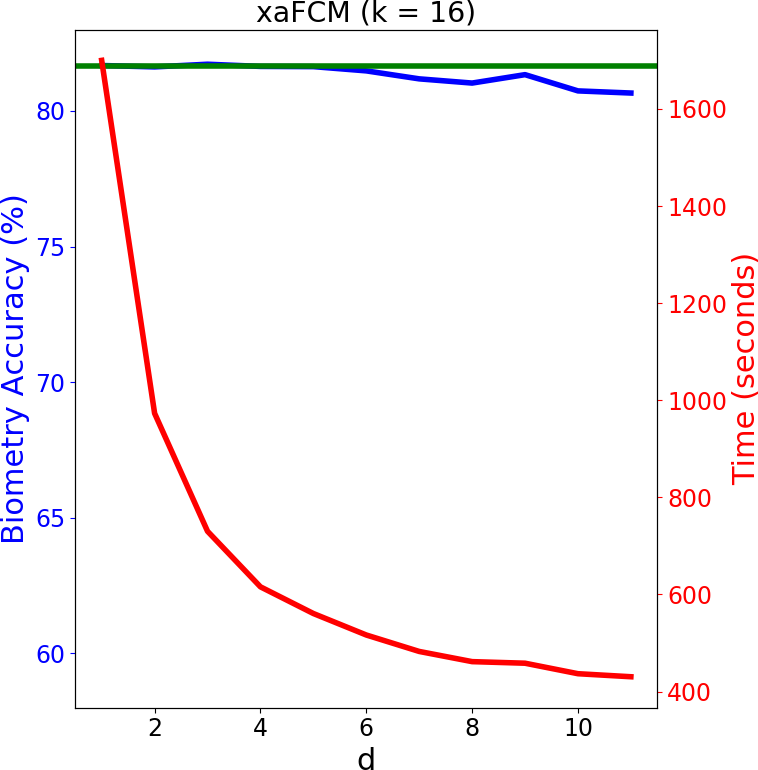}
	\end{minipage}
	
	\quad	
	
	\begin{minipage}[b]{0.4\textwidth}
		\includegraphics[width=\textwidth]{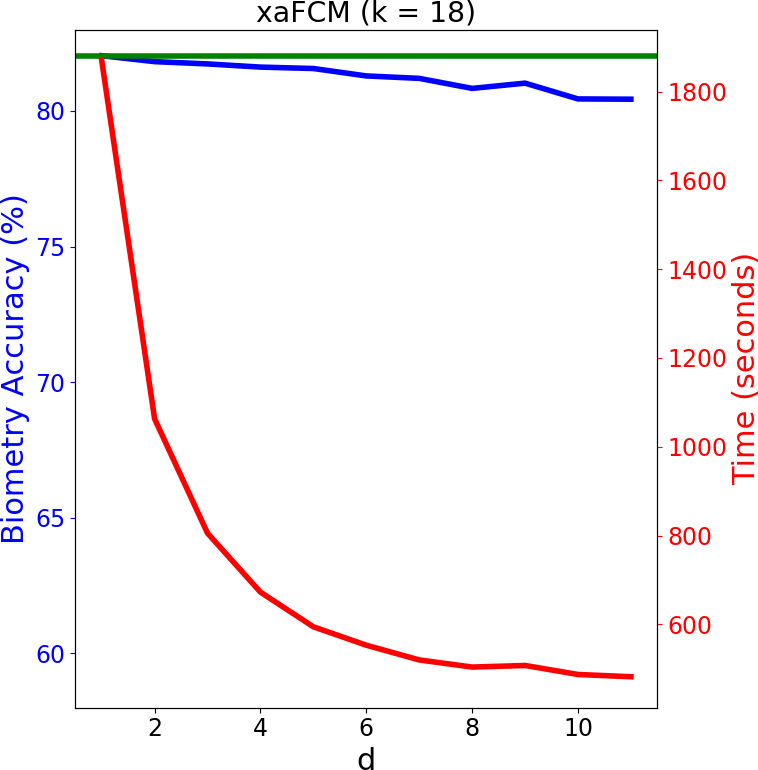}
	\end{minipage}
	\hfill
	\begin{minipage}[b]{0.4\textwidth}
		\includegraphics[width=\textwidth]{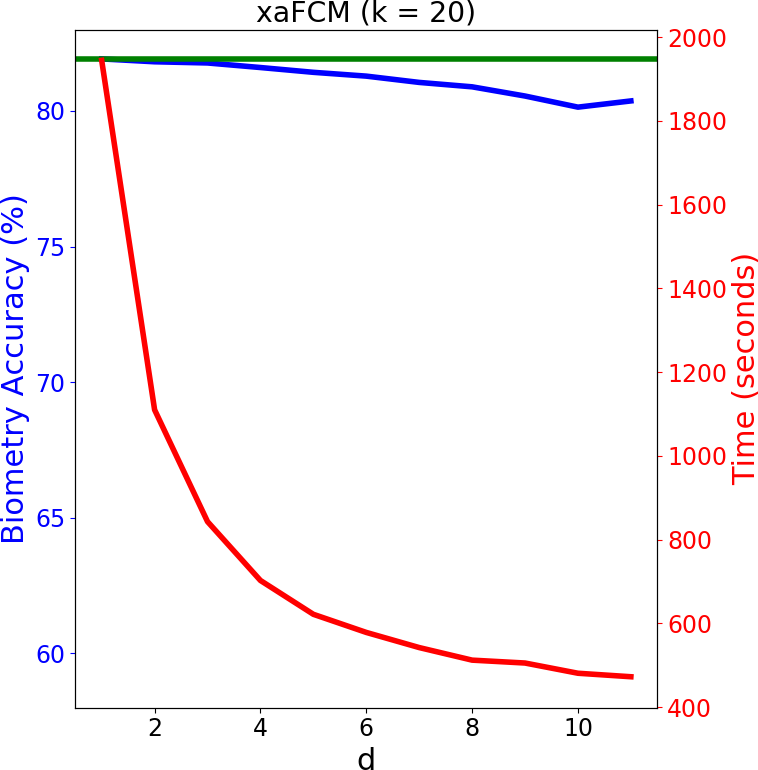}
	\end{minipage}
	\hfill
	
	\caption{xaFCM biometry process using high contexts $k$ (fixed); depth $d$ is changing from 1 to 11; the blue line represents the biometry accuracy (in \%); the red line represents the time (seconds); the green line represents the accuracy that a \textrm{FCM} with the same context would obtain.}\label{imgs_xafcm_biometry_and_time}
\end{figure}

The database used in our experiments was collected \textit{in house} \cite{Ferreira2016, carvalhoibpria2017}, where 25 participants were exposed to different external stimuli -- \textit{disgust}, \textit{fear} and \textit{neutral}. Data were collected on three different days (once per week), at the University of Aveiro, using a different stimulus per day.

The data signals were collected during 25 minutes on each day, giving a total of around 75 minutes of ECG signal per participant. Before being exposed to the stimuli, during the first 4 minutes of each data acquisition, the participants watched a movie with a beach sunset and an acoustic guitar soundtrack, and were instructed to try to relax as much as possible. 

By using a database where the participants were exposed to different stimuli, we can check if the emotional state of participants affects the biometric identification process.
The database is publicly available for download in \footnote{\url{http://sweet.ua.pt/ap/data/signals/Biometric_Emotion_Recognition.zip}}.

\begin{figure}[!htb]
	\centering
	\includegraphics[width=0.7\textwidth]{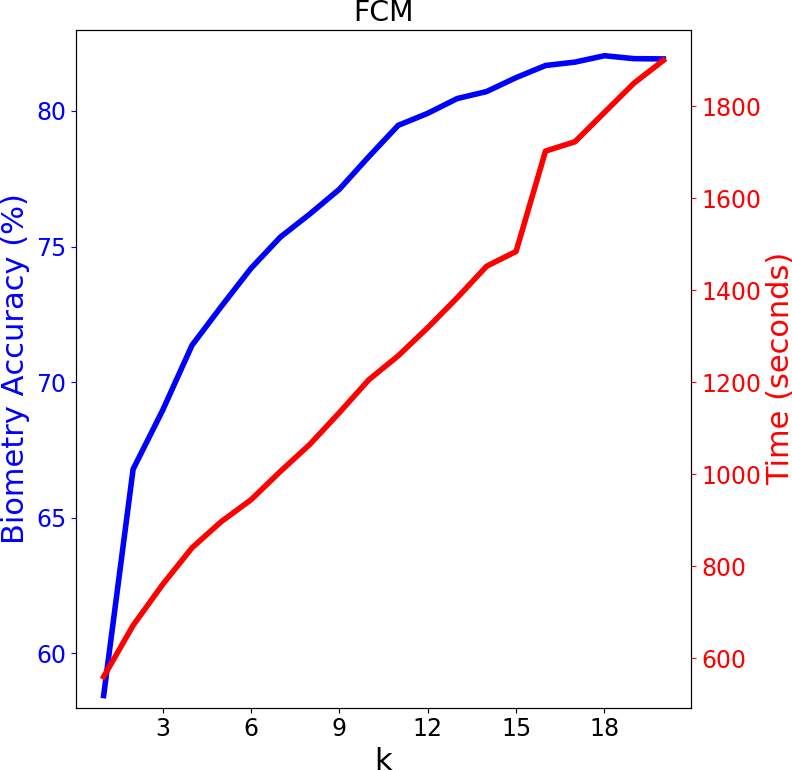}
	\caption{FCM biometry process: context $k$ changing; the blue line represents the biometry accuracy (in \%) and the red line represents the time of execution (seconds).}\label{img_biometry_fcm}
\end{figure}

After all the already explained preprocessing steps are complete, the process in which we perform the biometric identification is the following:

\begin{enumerate}
	\item Use the complete ECG signals from two days, in order to build a \textrm{xaFCM} model that describes each of the participants;
	\item For the remaining day, split the signal, such that each segment has 10 consecutive heartbeats inside it;
	\item ``Compress" (compute the \textrm{NRC}) each of the segments obtained in the previous step using each of the models obtained in the first step;
	\item The model which produces a lowest result is chosen as the candidate for biometric identification.
\end{enumerate}

The justification for the first step is that we do not want to use any information from the ECG of the day where we are trying to perform the ECG biometric identification, since, if we used that information, our results would not match a real situation.

The number of heartbeats needed for ECG biometric identification is undoubtedly useful when building a biometric identification system -- any system should ask participants to provide data for identification, using the smallest time interval that is possible, for practical reasons. Based on the results from a previous study \cite{carvalhoibpria2017}, we concluded that 10 heartbeats is a good trade-off between collection time (which should be as low as possible) and statistical relevance of the data.

\quad

All the experiments were implemented and ran using Python 3.5 (Linux 64 bits) on an Intel(R) Core(TM) i7-6700 CPU @ 3.40GHz, with 32GB of RAM. For simplicity of code, we have not parallelized the process yet -- therefore, only one logical core was used for each experiment.

In Fig.~\ref{img_biometry_fcm}, it is possible to see a plot with the accuracy obtained for the process described, by using \textrm{FCM} models, with all possible values of $k$ from 1 up to 20. In the red line, it is also possible to see how much time does this process take in total. An important fact is that the time taken to perform the biometry is approximately directly proportional to the size of the context, $k$, used.

Since the purpose of this paper is to show the appropriateness of \textrm{xaFCM} models, in Fig.~\ref{imgs_xafcm_biometry_and_time} are shown six examples of the same experiment, but instead of changing the context $k$, we have chosen a fixed value of $k$ and tested all possible values of $d$, the depth of the \textrm{xaFCM}s. From these plots, it is possible to see that the time taken to perform the biometry process for the whole database is up to 3-4 times shorter when using high values of $d$, having, usually, accuracy ratios comparable with the \textrm{FCM}s of the same order $k$.

\begin{figure}[!htb]
	\centering
	\includegraphics[width=0.8\textwidth]{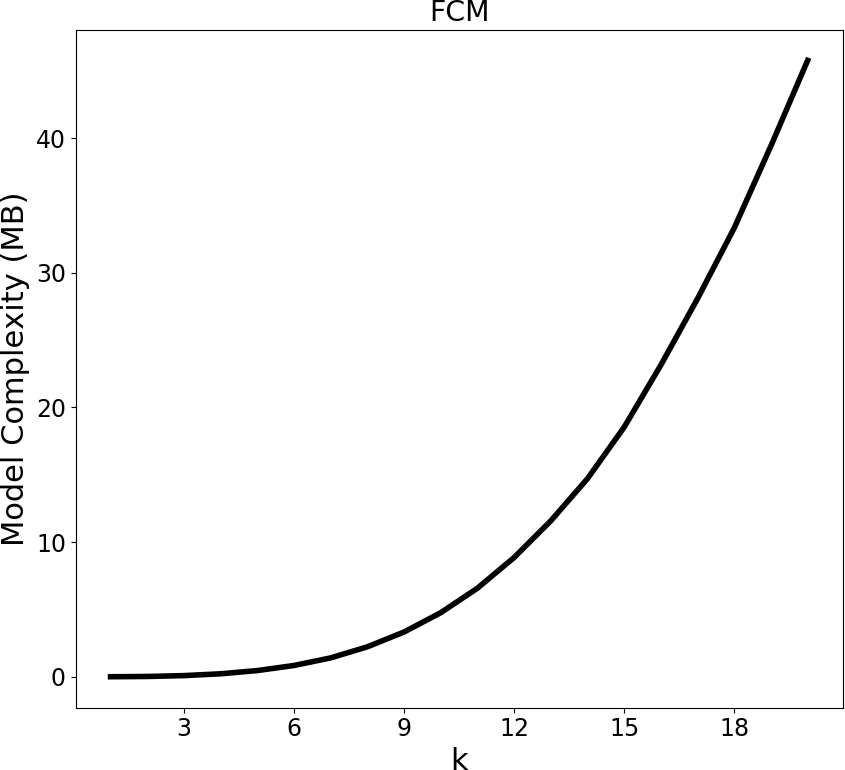}
	\caption{FCM average model complexity per participant - context $k$ changing.}\label{img_fcm_model_complexity}
\end{figure}

On the experiments using ``lower" values for the context $k$ (in this case, $k \leq 14$), it is possible to notice a minor improvement in terms of accuracy as the $d$ increases, at least for the first values of $d$ ($d \leq 7$, more or less). This makes us think that increasing the depth $d$ behaves in a similar way to increasing the depth $k$ of the \textrm{xaFCM}, without the additional cost in terms of testing speed (quite the opposite, actually) and the memory needed does not increase so much as it would by increasing $k$ (Fig.~\ref{img_fcm_model_complexity}).

In higher contexts $k$ we get the same advantages in terms of computing time and memory requirements, however, after a certain point, there is just no real benefit from increasing neither the context $k$, nor the depth $d$, since we are looking for ``too specific" patterns, that may not appear again on the segments being tested -- which, making an analogy to machine learning, we would be overfitting to the training data.

\begin{figure}[!htb]
	\centering
	\includegraphics[width=0.8\textwidth]{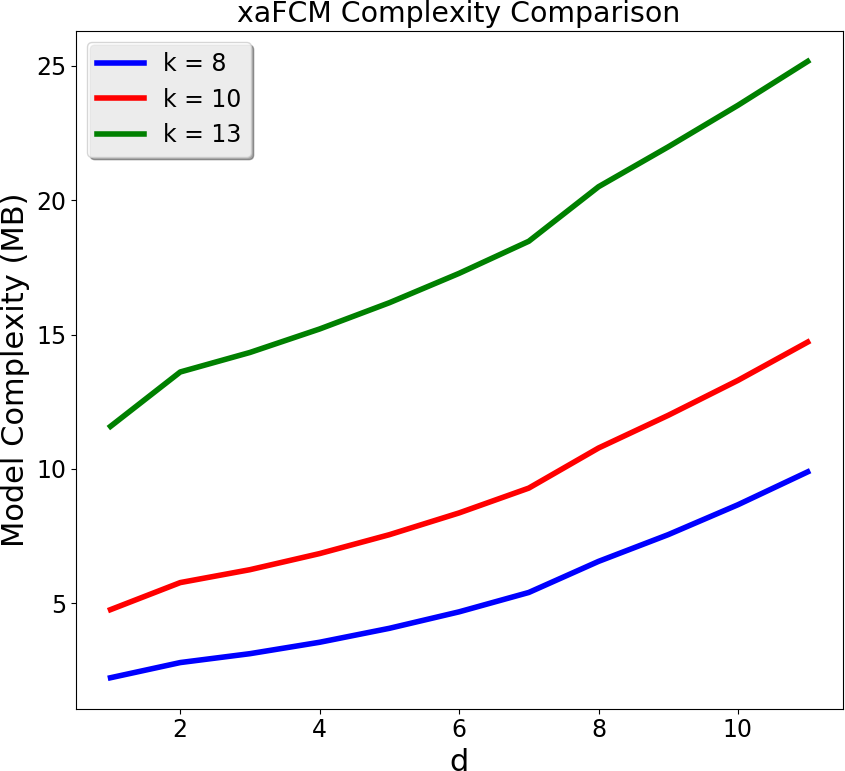}
	\caption{xaFCM average model complexity per participant - context $d$ is changing; $k$ is fixed.}\label{img_xafcm_model_complexity_comparison}	
\end{figure}

Another aspect we wanted to show, regarding the advantages of using \textrm{xaFCM}s, is the model complexity. In order for the biometric identification to be executed fast, in practice, it is needed to have all the participant models previously loaded into memory. This usually does not pose a problem, if there are not many participants, but it may be useful for building a real biometric identification system.

In Fig.~\ref{img_fcm_model_complexity}, we can see that by increasing the context $k$ of \textrm{FCM} models, the complexity of each model increases exponentially. From our interpretation, a way to avoid this exponential increase is to use an \textrm{xaFCM} with an order slightly lower and increase its depth $d$. In order to show this, we display the complexity of such models in Fig.~\ref{img_xafcm_model_complexity_comparison}.

\section{Application 2 -- DNA Sequence Relative Similarity}\label{section_dna}

\begin{table*}[]
	\centering
	\caption{CPU time and memory usage (RAM) of the experiments with DNA sequences.}
	\label{table_dna_results}
	
	\tiny
	
	\begin{tabular}{|c|c|c|c|c| }
		\hline
		Parameters & Average Time to & Average Time to  & Average Memory & Total Time to\\
		(context $k$ and depth $d$) 		  & Learn the Model     &   Compress       &   per model    &   run the experiment		\\\hline	
		
		$k = 12$, $d=1$    & 1649.6 sec & 1580.5 sec & 5043.2 MB &	274.4 hours	\\\hline		
		
		$k = 12$, $d=8$    & 2181.2 sec & 269.5 sec   & 14350.3 MB			& 59.5 hours \\\hline	
		
	\end{tabular}
\end{table*}
An approach for computing the similarity of a sequence relatively to other is to calculate the \textrm{NRC} using one of them as reference and the other as the target. In previous works, this has been done using \textrm{FCM} compressors \cite{Pinho2011, Pratas2014, Pratas2014a, Pratas2016a}. 

In order to show that the \textrm{xaFCM}s are also suitable for this application, we ran some simulations using the human and chimpanzee DNA sequences, removing the unknown symbols (N). The idea was to use each chromosome of the human species as reference and then compress each chromosome of chimpanzee as the target, using exclusively the model from the reference. Since we know from evolution theory that these two species are closely related \cite{dna_lineage}, it is expected that, when we are compressing homologous pairs of chromosomes, the \textrm{NRC} should be lower than on the other cases. 

\begin{figure*}[!htb]
	\begin{center}
		\includegraphics[width=0.99\textwidth]{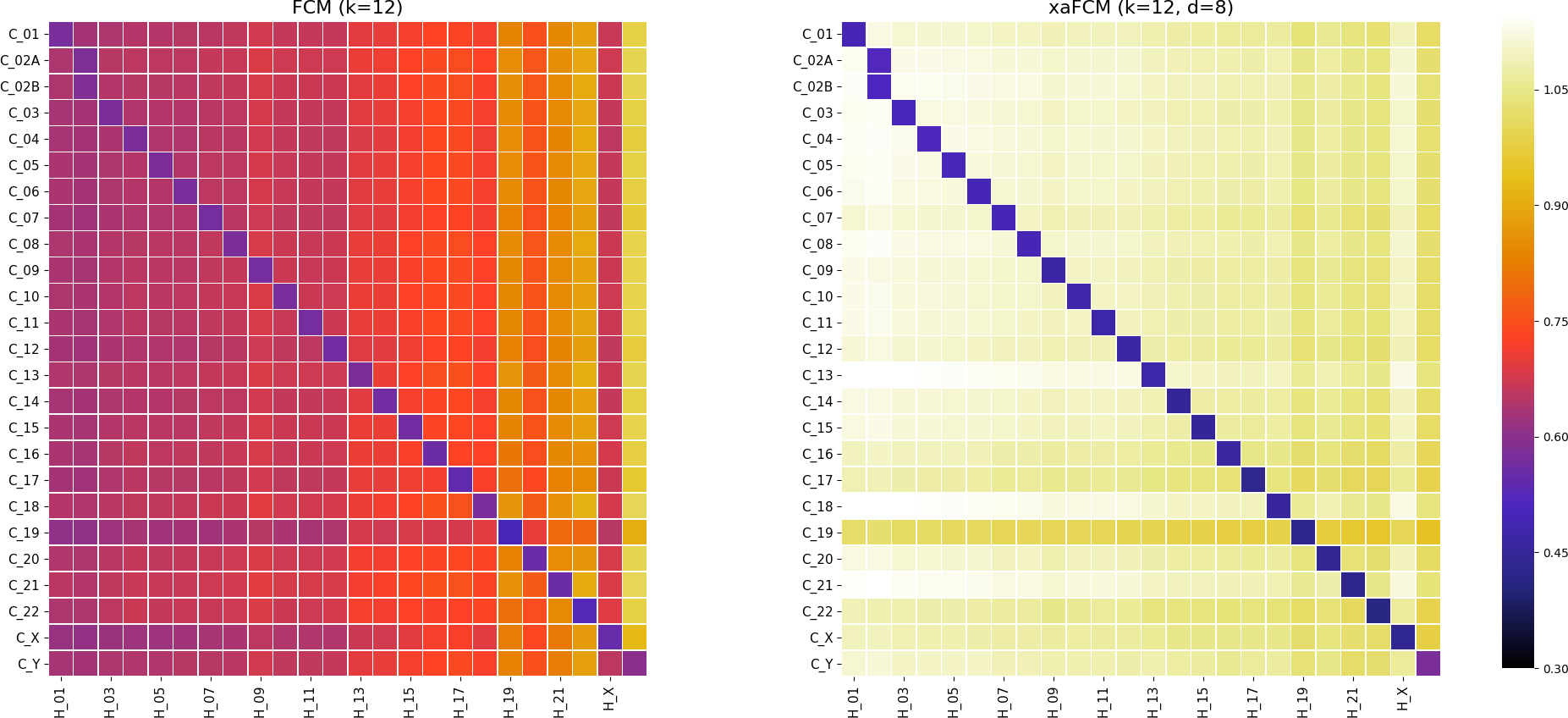}
	\end{center}
	\caption{Normalized compression of the chimpanzee (C) chromosomes relatively to the Human (H), using: \textbf{(left)} \textrm{FCM} ($k=12$); \textbf{(right)} \textrm{xaFCM} ($k=12$, $d=8$).}\label{img_dna_heatmaps}
\end{figure*}

\begin{figure*}[!htb]
	\centering
	\begin{minipage}[b]{0.97\textwidth}
		\includegraphics[width=\textwidth]{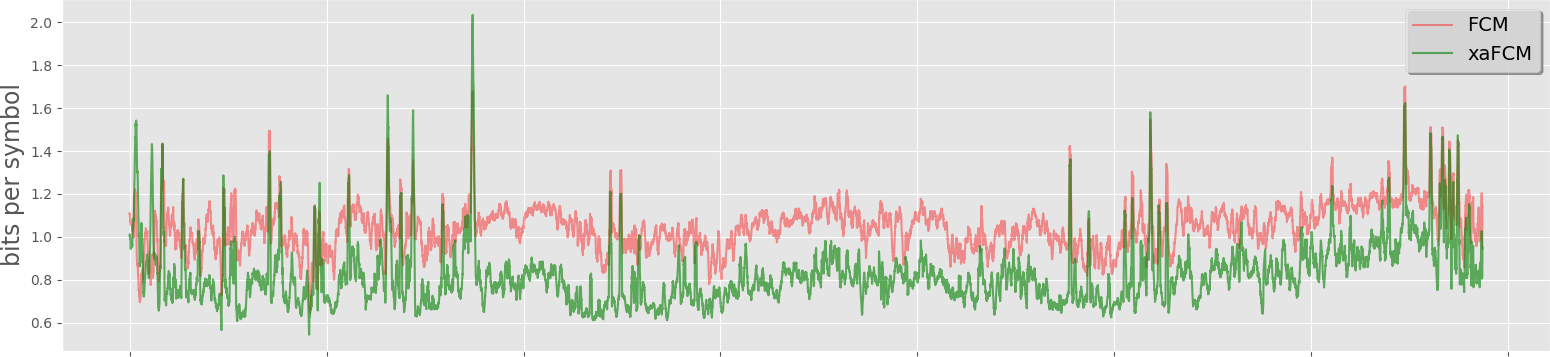}
	\end{minipage}	
	\hfill
	\caption{Profiles of information content of the chimpanzee chromosome 22 relatively to the human chromosome 22 using different models (\textrm{FCM} and \textrm{xaFCM}).}\label{img_dna_bits_per_symbol}
\end{figure*}

To perform the experiment, we used the assembled human chromosomes 1 to 22, X and Y (3.1GB of data in total) and assembled chimpanzee chromosomes 1, 2a, 2b, 3 to 22, X and Y (3.2GB of data in total)\footnote{All the assembled genome data were downloaded from ftp://ftp.ncbi.nlm.nih.gov/genomes/}. We ran two different simulations: the first one, with a \textrm{FCM} of context $k=12$; the other with a \textrm{xaFCM} with $k=12$ and $d=8$. All the experiments ran on a server with 16-cores 2.13GHz Intel Xeon CPU E7320 and 256GB of RAM, but the implementation used a single core.

Table \ref{table_dna_results} shows the average times taken by each experiment, as well as the average memory needed to store the \textrm{xaFCM} model to represent the human chromosomes.

It is clear from these results that the \textrm{xaFCM}s are almost $d$ times faster than an \textrm{FCM} of the same order $k$. Another advantage is that the memory needed for the \textrm{xaFCM}s does not increase exponentially with $d$.

The \textrm{NRC} results for the two simulations, with $k=12$, can be seen in Fig.~\ref{img_dna_heatmaps}.

It is possible to notice that the heatmap corresponding to the \textrm{FCM} shows better compressions on average. 
However, using the ``perfect" relative compressor, we would expect the \textrm{NRC}s to be as low as possible on the diagonal of the matrix"\footnote{Not exactly the diagonal, because of the second chromosome of the chimpanzee is split into 2a and 2b, making the matrix not a square one.}, since they represent related chromosomes. The other squares should have higher \textrm{NRC}s, as they have more variation. This is exactly what happens on the \textrm{xaFCM} test (bottom one in Fig.~\ref{img_dna_heatmaps}).

This becomes even more clear when we are comparing the compression along the same sequence, as can be seen in Fig.~\ref{img_dna_bits_per_symbol}.

\section{Conclusions and Future Work}

We have shown that \textrm{xaFCM}s are good candidates to represent models for ECG biometric identification. When compared with \textrm{FCM}s, with the same memory usage, better accuracy ratios are usually obtained, using up to around 3-4 times less time to compute the \textrm{NRC}s (depending on the choice of $d$).

The gains in computational speed increase in the DNA sequence given the higher order of data length. Our experiments show that it is possible to use them for DNA sequence pattern recognition, making them a suitable alternative to the traditional \textrm{FCM}s.

These are promising results, and it seems appropriate to infer that the \textrm{xaFCM}s can be suitable to some other applications, specially when the problem of memory usage or testing speed are crucial. For that reason, in the near future, we plan to test them in different applications, where \textrm{FCM}s have proven suitable, like image pattern recognition \cite{Pinho2011a, Pinho2011b, Pinho2014b} and authorship attribution \cite{Pinho2016a}.

\section{Acknowledgments}
This work was partially supported by national funds through the FCT - Foundation for Science and Technology, and by european funds through FEDER, under the COMPETE 2020 and Portugal 2020 programs, in the context of the projects UID/CEC/00127/2013 and PTDC/EEI-SII/6608/2014. S. Brás acknowledges the Postdoc Grant from FCT, ref. SFRH/BPD/92342/2013.

\bibliographystyle{acm}
\bibliography{library}

\end{document}